\documentclass[paper]{JHEP3}
\usepackage{epsfig}

\newcommand{\be}{\begin{equation}}
\newcommand{\ee}{\end{equation}}
\newcommand{\nnb}{\nonumber}
\newcommand{\bea}{\begin{eqnarray}}
\newcommand{\eea}{\end{eqnarray}}
\newcommand{\bean}{\begin{eqnarray*}}
\newcommand{\eean}{\end{eqnarray*}}
\newcommand{\eps}{\varepsilon}

\newcommand{\half}{\frac{1}{2}}

\newcommand{\bvec}[1]{\vec{#1}}

\newcommand{\AME}{{\tt AMEGIC++}}
\newcommand{\APA}{{\tt APACIC++}}

\newcommand{\St}{\tilde{S}}

\title{Implementing initial state radiation for lepton 
       induced processes in \AME}
 
\author{A. Sch{\"a}licke$^{a,b}$, F.\ Krauss$^b$, 
        R.\ Kuhn$^{a}$, and G.\ Soff$^a$\\
$^a$Institut f\"ur Theoretische Physik, TU Dresden, 01062 Dresden,
Germany\\
$^b$Cavendish Laboratory, University of Cambridge, Cambridge CB3 0HE,
U.K.\\
E-mail: \email{schalicke@hep.phy.cam.ac.uk},
        \email{krauss@hep.phy.cam.ac.uk},
        \email{kuhn@theory.phy.tu-dresden.de}
}

\abstract{
We have implemented the method of Yennie, Frautschi, and Suura up to 
first order in $\alpha$ for the simulation of QED Initial State 
Radiation in lepton induced processes in \AME. We consider $s$--channel 
processes via the exchange of scalar or vector resonances at electron 
and muon colliders.}
 
\keywords{Standard Model, Supersymmetric Standard Model, Higgs Physics, 
          Electromagnetic Processes and Properties}
 
\preprint{Cavendish-HEP-02/02}

\begin{document}

\section{Introduction}\label{c_intro}

Initial State Radiation (ISR) is the most important QED correction 
to the Born cross section. For instance, in the case of electron 
positron annihilations at energies beyond the $Z$--threshold 
it leads to the so--called radiative return causing a hugely increased
cross section. There, via multiple emission of photons, the electron 
positron pair tends to reside on the $Z$--resonance (returning on 
it) picking up the propagator in its resonance region. In this way, 
the correction term becomes several times as large as the Born term
at the initial center of mass energy. Evidently, the effect of ISR 
is of great significance for the interpretation of experimental 
results. However, since initial state photons tend to be collinear 
and soft they easily escape detection. In the detector, they 
tend to leave only imprints by a reduced overall energy and some netto 
transverse momentum and -- possibly -- by a boost of the visible 
final state along the beam axis. Due to this rather indirect method 
of measuring it, the precise simulation of ISR is crucial for most 
experimental analyses. 

Several strategies exist for this task:
\begin{enumerate}
\item The structure function approach

  The idea behind this approach is to simulate ISR by a 
  probability density to find an electron with reduced momentum 
  inside an incoming electron. Accordingly, the missing momentum 
  is then attributed to photon radiation. That way the total 
  cross section $\sigma_{\rm tot}$ including ISR is defined as 
  the leading order cross section $\sigma_{\rm B}$ convoluted 
  with two structure functions $\Gamma_{ee}$:
  \be
    \sigma_{\rm tot} =  \int dx_1 dx_2 \Gamma_{ee}(x_1,s) 
                                       \Gamma_{ee}(x_2,s) 
                        \sigma_{\rm B}(x_1 x_2 s)
                        \;.
  \ee
  Within the structure functions logarithmically enhanced
  contributions are exponentiated and thus re-summed to all orders with
  additional terms covering non--factorisable QED effects up to a
  finite order $\alpha^n$. Additionally, terms describing weak and
  strong effects can be taken into account. 
  However, since the structure function is integrated over all photon  
  energies and transverse momenta, it is limited to situations in 
  which only the overall effects of energy reduction and longitudinal
  boost via ISR are of interest. Arbitrary phase space cuts are
  difficult to implement. Specific photonic observables like for
  instance the number of photons above some energy threshold or their
  transverse momenta are beyond the reach of this approach.
\item The parton shower approach
    
  In this approach, the electron radiation is followed step by 
  step from the comparably low scales of the order of the electron 
  mass connected to the incoming electrons to the high energy scale 
  of the actual hard process. After setting up an electron 
  distribution function at the initial scale, any individual 
  photon emission is governed by the well--known splitting function 
  \be
    P_{ee}(z) = \frac{\alpha}{2\pi}\,\frac{1+z^2}{1-z}
  \ee
  leading to a Dokshitzer--Gribov--Lipatov--Altarelli--Parisi 
  (DGLAP) evolution equation for the structure function. Technically, 
  the original structure function is being reproduced statistically at
  each scale via backward evolution and re-weighting, where the
  analytic integration is replaced by Monte Carlo integration.  
  Apparently, photons are generated explicitly which allows for instance
  the application of arbitrary phase space cuts.
  However, one of the weaknesses of this approach is that it re-sums 
  only the logarithmic enhanced contributions in the DGLAP equation. 
  Even though the radiation of the hardest photon can be corrected
  with the first order matrix element,
  it is not clear how to systematically improve to higher
  perturbative order, i.e.\ how to 
  make sure that matrix element expressions for the emission of 
  additional photons are recovered for arbitrary processes and 
  arbitrary numbers of photons.
\item The approach by Yennie, Frautschi and Suura (YFS) \cite{YFS}

  Within this approach the explicit generation of photons can  
  be corrected systematically to all orders in the coupling 
  constant. The basic idea here is to introduce some arbitrary 
  infrared cut--off on the photon energies and treat low energy 
  real photons as ``un-resolvable''. Their respective 
  contribution will cancel the emerging virtual infrared divergences 
  to all orders leading to finite factorisable terms which can be
  exponentiated. In this framework correction terms for harder photons
  can be easily introduced. Since this is the approach we are 
  going to follow we postpone a more thorough discussion.
\end{enumerate}

In this paper we want to give a description of the YFS method implemented
in \AME\ \cite{amegic}. The outline is as follows: 
In section \ref{c_method}, after briefly reviewing the YFS approach and 
a short introduction to the nomenclature, we will introduce our new
approach to calculate matrix element corrections, where we make use of
the ability of \AME\ to calculate (almost) arbitrary tree level matrix
elements automatically. In fact, it can be anticipated that our method
might allow for the construction of a more generic simulation code for
QED initial state radiation in arbitrary processes in the near future.
In section \ref{c_implementation} we will discuss the actual implementation 
of our approach and the interplay of ISR with the calculations of
matrix elements and cross sections. We justify our method in section
\ref{results} by confronting our results for the radiative return to
the $Z$-pole with experimental data from the LEP experiment. The
generality of our program is illustrated by the application to the
$s$--channel production of Higgs bosons at a possible muon
collider. We will conclude with summarizing remarks.

\section{Method}\label{c_method}

In this section we want to discuss our new method to include the 
effect of exact matrix elements in the generation process of initial 
state photons along the lines of the YFS approach. We will begin with a 
mini--review of this method suitable for Monte Carlo implementation.

\subsection*{Mini-review of the YFS approach}
The idea underlying the YFS 
approach is to separate the phase space for the emission of real photons 
into two regions via a cut--off $\epsilon$ on the energy fractions, 
such that photons are coined infrared if their energies 
$\omega_\gamma < \epsilon E_{\rm beam}$. 
The contribution of these infrared photons is then used to cancel the 
virtual infrared divergences order by order in $\alpha$. The remainders
of this procedure factorize and can be exponentiated into an universal 
factor, the YFS form factor: 
\bea
 F^{\rm YFS}(\eps)
 &=& \exp\left[ 2\alpha (B + \tilde B)\right]\nnb\\
 &=& \exp\left[ 
  \frac{\alpha}{\pi} 
  \left(\half\ln\frac{s}{{m}^2} -1 + \frac{\pi^2}{3} \right)
  \,+\,
  \frac{2\alpha}{\pi}
  \left(\ln\frac{s}{{m}^2}-1\right)\ln\eps  
  \right] 
  \;.
\eea
The virtual part is given by
\be
 \alpha B = \int \frac{ d^Dk}{(2\pi)^D}  \frac{1}{k^2}  S(k) 
\ee
whereas the real contribution reads 
\be
\label{f_B_tilde}
 \alpha \tilde B = \int\limits_{\omega < \epsilon E} 
                     \frac{ d^{D-1}k}{(2\pi)^{D-1} 2\omega}  \St(k)\,.
\ee
The functions $ S(k)$ and $ \St(k)$ denote the well known universal
factorizing  ``radiation factors'' or ``eikonal factors'' for virtual
and real photons for a pair of two external charged lines with
four--momenta $p_1$ and $p_2$  
\bea
\label{f_s_k}
 S(k) &=&
  - \frac{\alpha}{8\pi} 
    \left[  \frac{2p_1^\mu - k^\mu}{k^2 - 2kp_1} 
          - \frac{2p_2^\mu - k^\mu}{k^2 - 2kp_2} \right]^2 \;, \nnb\\
 \St(k) &=& 
   \frac{\alpha}{4\pi} \,  
    \left[  \frac{p^\mu_2}{kp_2} 
          - \frac{p^\mu_1}{kp_1} \right]^2\,.
\eea
The emission of the visible real photons can then be corrected 
systematically to all orders in $\alpha$ to reproduce exact results as 
given by the corresponding matrix elements. To be more specific, let us
consider as an example $2\to2$ processes of the type 
$l^+(p_1)\, l^-(p_2) \to f(q_1)\, \bar f(q_2)$, where the $p_i$ and $q_j$
label the four--momenta of the external particles. The total cross 
section including arbitrary numbers of real or virtual photons can
be written as
\bea\label{master_f}
  \sigma 
  & = & 
    F^{\rm YFS}(\eps)\;
    \int\frac{d^3 \bvec{q}_1}{q_1^0}\frac{d^3 \bvec{q}_2}{q_2^0}\; 
    \sum_{n=0}^\infty \left\{  \frac{1}{n!} 
    \left[ \prod_{i=1}^n \frac{d^3 \bvec{k}_i}{k_i^0}\,
    \St(p_1,p_2,k_i)
    \;\Theta\!\!\left(\frac{2 k_i^0}{\sqrt{s}}-\eps
     \vphantom{\sum_{j,l=1, j\ne l}^n}
    \right) \right]\right.  \nnb\\
  && \hspace*{4.9cm} 
     \delta^4 \left(p_1 + p_2 -q_1  -q_2 - \sum_{i=1}^n k_i \right)
     \nnb\\
  && \hspace*{4.8cm} \left.
     \left( \beta_0 + 
                  \sum_{j=1}^n{\frac{\beta_1(k_j)}{\St(k_j)}} +
                  \sum_{j,l=1, j\ne l}^n
                              {\frac{\beta_2(k_j,k_l)}
                                     {\St(k_j)\St(k_l)}} + 
           \dots \right)\right\}\;.\nnb\\ 
\eea
The YFS form factor $F^{\rm YFS}(\eps)$ covers the contribution of
factorizing soft real and virtual photons to all orders. The integral
over the phase space of the final state particles consists of the
integral over the two outgoing momenta $q_1$ and $q_2$ plus a sum over
all possible numbers of photons $k_i$ with energy fractions above the
resolution threshold $\epsilon$. This constraint is reflected in the
$\Theta$-functions. The conservation of total four--momentum is
enforced by the $\delta$-function in the second line. Finally, the
last line includes the matrix element corrections. The $\beta_n$
denote infrared safe combinations of cross sections and eikonal
factors with $n$ additional real photons as well as all finite
contributions of any number of virtual photon loops. In practice,
the order of virtual contributions is limited. In $\beta_n^{(l)}$ the
superscript $l$ denotes the total number of virtual and real photons
taken into account, i.e.\ the order of perturbation theory. From there
we can read off the number of virtual photons, given by $(l-n)$. 

For a treatment exact up to first order $\alpha$, it is sufficient to 
determine $\beta_0^{(0)}$, $\beta_0^{(1)}$ and $\beta_1^{(1)}$,
i.e.\ the cross sections
for $l^+(p_1) \, l^-(p_2) \to f(q_1)\, \bar f(q_2)$ up to one loop and
for $l^+(p_1) \, l^-(p_2) \to f(q_1) \,\bar f(q_2) \, \gamma(k_i)$ at
the tree level. They are defined as 
\bea\label{defbeta}
\beta_0^{(0)}       
    &=& \rho_0^{(0)}\;,\qquad 
\beta_0^{(1)}       
    = \rho_0^{(1)}\;,\nnb\\
\beta_1^{(1)}(k_i)  
    &=& \rho_1^{(1)} (k_i) - \St(k_i)\,\beta_0^{(0)}\;.
\eea
The $\rho_n^{(l)}$ are the fully differential cross sections for the hard 
process $l^+(p_1)\, l^-(p_2) \to f(q_1) \,\bar f(q_2)$ with $n$ real photons
added and up to $l$th order in $\alpha$, and $\St$ denotes the eikonal factor
as defined in Eq.(\ref{f_s_k}). 

Analyzing the master equation, Eq.(\ref{master_f}), we realize that 
\begin{enumerate}
\item the number of photons that are explicitly produced is not constrained,
\item the resolved photons are allowed to have transverse momentum,
\item the cross section is independent of $\epsilon$ (in fact, only the 
      number of explicitly produced photons depends
      on the resolution scale), and
\item that in contrast to the parton shower approach any photon emission
      is corrected by the appropriate matrix element(s) to the chosen
      order in $\alpha$ including interference effects in multi photon
      final states.  
\end{enumerate}

\subsection*{The MC algorithm}
A Monte Carlo implementation for the YFS approach outlined above 
was described in \cite{yfs2} for the first time. The proposed procedure
there was 
\begin{enumerate}
\item to chose the c.m.\ energy with the form factor $F^{\rm YFS}$,
\item to determine the number of resolvable photons,
\item to supply each of the photons with a four--momentum $k$ according to the
      eikonal factor $\St(k)$,
\item to reject all momenta outside their kinematical limits, and
\item to correct the configuration chosen with an additional weight
      stemming from the exact matrix element.
\end{enumerate}

\noindent
A more detailed description of the Monte Carlo steps can be found 
in Appendix \ref{a_mc_alg}.

\subsection*{Implementing matrix element corrections}
The corrections on the photon distributions and the cross sections due
to exact matrix elements can be read off from  Eq.(\ref{master_f}).
Up to first order in $\alpha$ the corresponding weight reads:
\bea
  w_{\rm ME}  =     \frac{1}{\sigma_B(s')}
  \biggl\{  \beta^{(1)}_0(p_i,q_j) + 
   \sum_{n=1}^{N_\gamma} \frac{\beta_1^{(1)}(p_i,q_j,k_n)}
                         {\St(p_i,k_n)} \biggr\}\;,
\eea
where again the $p_i$ are the incoming momenta, the $q_j$ are the outgoing 
momenta and the $k_n$ are the momenta of the explicitly generated photons. 
In the following we will again restrict ourselves to the description 
of the process
$e^+ e^- \to f \bar f$, i.e.\ a process with two incoming and 
two outgoing momenta only. However, the extension to more final state
particle is straightforward.

An implementation of this weight might seem trivial, however, one
serious question remains: How do we calculate matrix elements in
situations, where the number of generated photons exceeds the number
incorporated in the matrix element? Matrix elements are only
unambiguously defined if all incorporated momenta comply with
four--momentum conservation. This leads to a problem in cases when 
some of the generated photon momenta are omitted.  

In general, it is solved by applying an unique description of how to
evaluate $\beta$ functions. Two approaches have been developed in the
past to deal with this situation, namely the projection method
\cite{yfs2} and the extrapolation method \cite{koralz}. In the
following we will briefly discuss both methods in order to point out
their advantages and disadvantages. Our own solution will be presented
afterwards. 

The projection method defines for each $\beta_n$ function a 
mapping of the four--momenta from the full phase space with $N_\gamma$  
photons to the restricted phase space with only $n$ photons. Therefore,
$N_\gamma-n$ momenta are completely discarded. The four--momenta of the 
remaining particles are transformed such that they again comply with
four--momentum conservation. In general, this is achieved by a
sequence of scalings and boosts applied in individual combinations on
the incoming and outgoing particles. For instance, for the leading
order term $\beta_0$ the projected momenta $p'_i, q'_j$ are obtained
by boosting the incoming momenta $p_1$ and $p_2$ into their common
c.m.\ frame and scaling them to fulfill the condition 
$(p'_1 + p'_2)^2 = (q_1 + q_2)^2 = s'$. The outgoing momenta $q'_1$
and $q'_2$ are thus defined by boosting the original $q_j$ into their
common c.m.\ frame. A projection for higher order corrections is more
involved and will be discussed at the end of this section in the
framework of our new approach. 

The extrapolation method, on the other hand, defines the matrix
element in any phase space with additional photons, i.e.\ the momenta
are kept untouched but the matrix element is rewritten in a suitable
set of variables like e.g.\ angles and energies. Obviously, there is
no universal way to construct this transformation and therefore this
has to be done for each matrix element separately. 

As an example we briefly review the corresponding algorithm in the
program KoralZ, that is dealing with processes of the type $e^+ e^-
\to f \bar f$ with $f\ne e^-$. The leading order term $\beta_0$ is
written as  
\be
\beta_0( p_i, q_j; k_1, \dots, k_{N_\gamma}) =
\half\left[ 
   \frac{d \sigma_0}{d \Omega}(s',\cos\vartheta_1)
 + \frac{d \sigma_0}{d \Omega}(s',\cos\vartheta_2) 
     \right]\;,
\ee
where
\be
\cos \vartheta_1 = \frac{\bvec{p}_1 \bvec{q}_1}{|\bvec{p}_1|| \bvec{q}_1|}  
  = - \frac{\bvec{p}_1 \bvec{q}_2}{|\bvec{p}_1|| \bvec{q}_2|}
\ee
and
\be
\cos \vartheta_2 = \frac{\bvec{p}_2 \bvec{q}_2}{|\bvec{p}_2|| \bvec{q}_2|} 
    = - \frac{\bvec{p}_2 \bvec{q}_2}{|\bvec{p}_2|| \bvec{q}_2|}
\;
\ee
are the angles between equal and opposite charged particles,
respectively. In the presence of additional photons these angles
differ, but $\beta_0$ is well calculable for any number of photons,
since the variables $s'=(q_1+q_2)^2$, $\cos \vartheta_1$, and 
$\cos\vartheta_2$ are defined in all cases. For the calculation of
$\beta_1$ the variable $s=(p_1 + p_2)^2$ and the polar angle of the
photon $\cos \vartheta_{\gamma}$ are used in addition to 
$\cos\vartheta_1$, $\cos \vartheta_2$ and $s'$.

However, any approach to define matrix element corrections has to
comply with two conditions:
\begin{itemize}
\item 
  First, the method has to guarantee that the original 
  result is retained in case the neglected ``spectator'' photons 
  have total zero four--momentum. In particular, this must hold 
  true in the case of zero spectators.
\item 
  Second, the procedure should not give rise to new large 
  logarithms at any order beyond the perturbative order up to
  which the weight is defined.
\end{itemize}

At this stage we would like to present a new method to define the 
$\beta$ functions. We aim at a general approach that takes full
advantage of the automatic calculation of tree-level matrix elements 
in \AME. We do not want to interfere with the automatic calculation 
as such, which relies on the construction of four--momenta and their 
explicit conservation. Hence, the method of choice is rather a 
realization of the projection method.
 
Remember that the first order infrared finite correction term is 
\be
  \beta_1(p_i,q_j,k_n)
  =  \rho_1^{(1)}(p_i,q_j,k_n) 
           - {\St}(p_i,k_n)\,\beta_0^{(0)}(p_i,q_j) 
  \;.\nnb
 \label{betaone}
\ee
For this correction term the projection method is well defined as long
as there is exactly one additional photon. Then we can apply the PM as
introduced for the leading order term $\beta_0$ by redefining
$\beta_1$ as 
\be
  \beta_1(p_i,q_j,k_n)
  =  \rho_1^{(1)}(p_i,q_j,k_n) 
           - {\St}(p_i,k_n)\,\beta_0^{(0)}(p'_i,q'_j) 
  \;.\nnb
\ee

For more than one photon the arguments of the differential cross section
$\rho_1^{(1)}$ have to be projected as well. Naively, one might want to apply 
the same technique as for $\beta_0$ with the photon $k_i$ included in the
outgoing momenta, but it turns out that this is not sufficient to describe
events with additional hard photons.

Our solution to this problem is the following:
\begin{enumerate}
  \item
    The second term, consisting of the eikonal factor multiplied
    with the Born term cross section is still treated as described above.
    Here, we observe that
  \begin{itemize}
    \item the scale of the evaluation of the eikonal is $s$, since 
      it is evaluated with the original vectors $p_1$ and $p_2$. This
      means that the effect of all other photons is ignored which
      is yields the correct behavior in order to cancel the eikonal 
      factors in the YFS exponent.
    \item
      The scale of the Born cross section, i.e.\ the scale
      of the $s$--channel propagator, is $s'$.
  \end{itemize}
  \item
    In the first term we have to obtain a similar structure. This 
    is achieved by the following steps:
    \begin{itemize}
      \item
        To calculate the photon part at the scale $s$, it is necessary
        to leave the incoming momenta as well as the photon 
        momentum untouched, but the outgoing momenta
        have to be modified. The new momenta $\hat q_1$ and 
        $\hat q_2$ are fixed by the constraint 
        \bean
            p_1 + p_2  &=& \hat q_1 + \hat q_2 + k_n\;.
        \eean
        Thus, the differential cross section with one photon is
        given by
        \be
             \rho_1^{(1)}(p_1,p_2,\hat q_1,\hat q_2,k_n) \;. 
        \ee
      \item
        In the previous step, the scale of the $s$--channel propagator
        has been changed to $\hat s = (\hat q_1 +\hat q_2)^2$ which
        can modify the cross section quite drastically.
        We cure this by a correction factor
        \be
          C = \frac{ \beta_0 (p'_1, p'_2, q'_1,  q'_2)}
              {\beta_0 (\hat p_1, \hat p_2,\hat q_1, \hat q_2)}\;.
        \ee
        Thus, the right peak structure is 
        restored. The new incoming momenta are naturally defined
        via $\hat s = (\hat p_1 +\hat p_2)^2$.
    \end{itemize}
\end{enumerate}
Altogether, $\beta_1$ is defined as
\be
 \beta_1( p_i, q_j,k_n)  =
 \frac{ \beta_0 (p'_i,  q'_j)}{ \beta_0 (\hat q_i, \hat q_j)}
  \rho_1 (p_i,\hat q_j, k_n) -
 \St(p_i,k_n) \,\beta_0 (p'_i, q'_j)\;.
\ee 

\section{Implementation}\label{c_implementation}
In this section we want to discuss some aspects of the interplay of \AME\ 
with its ISR module. A more thorough description of \AME\ is available in
the manual \cite{amegic}, here we just list the basic steps to calculate
cross sections:
\begin{enumerate}
\item As a first step, \AME\ sets up the model in which the calculation 
      is going to take place. This includes the definition of particles,
      the initialization of Feynman rules in terms of vertices and --
      possibly -- the determination of the widths of unstable particles.
      This last step already relies on the construction and evaluation of
      corresponding Feynman diagrams and their integration.
\item The construction of Feynman diagrams is performed in three steps:
      \begin{itemize}
      \item The construction of ``empty'' topologies, i.e.\ without
            the specification of lines and vertices,
      \item their mapping on ``filled'' Feynman diagrams with specified 
            propagators and vertices, 
      \item and, finally, their translation into helicity amplitudes which
            can be stored in library files.
      \end{itemize} 
\item The integration of the diagrams is done with help of an
      adaptive multichannel method \cite{multichannel}. The channels
      are constructed by inspection of the kinematical structure of
      corresponding Feynman diagrams.
\end{enumerate}
Implementing ISR in the framework of YFS affects the last two steps.
First, \AME\ has to provide one additional matrix element to calculate
the real photon matrix element corrections. The second step is
therfore extended as follows: 
\begin{itemize}
\item 
  An ``ISR--Photon'' is defined, i.e.\ an additional particle type
  is introduced, with all properties of an ordinary photon but
  exclusively coupling to the charged particles in the initial state. 
\item
  One ``ISR--Photon'' is added to the list of outgoing particles.
  Correspondingly, ``empty'' topologies with one additional leg are
  constructed. 
\item
  The Feynman diagrams for this additional process are generated and 
  translated into helicity amplitudes as pointed out above.
\end{itemize}

\noindent
During the integration, step 3, the following changes take place:
\begin{itemize}
\item 
  The reduced c.m.\ energy $\sqrt{s'}$ is determined, and a
  number of ISR photons is generated along the lines of the
  algorithm in Appendix \ref{a_mc_alg}.
  With this energy a set of incoming momenta is defined 
  in their c.m.\ frame.
\item
  The momenta of the final state are generated and the 
  leading order cross section is calculated with a multichannel method.  
  So far the effect of ISR is only manifest by the reduced c.m.\ energy.
\item
  The ISR weight is calculated using the generated final state momenta.
  Therefore, the matrix element prepared in step 2 with 
  one additional photon as well as the Born matrix element are used
  according to our method introduced in section \ref{c_method}.
  The obtained weight is combined with the phase space weight.
\item
  Finally, a boost of all incoming and outgoing momenta is performed 
  in a way that
  all momenta, i.e.\ including the ISR photons, comply with
  four--momentum conservation. Note that this step does not
  modify the result of the integrated cross section but it is
  mandatory, when \AME\ is used inside a Monte Carlo generator.
\end{itemize}

\section{Results}\label{results}
To demonstrate the quality of our approach we will discuss two examples:
\begin{itemize}
\item Processes of the type $e^+e^- \to \gamma^*,Z^* \to f\bar f$ at
      LEP2 energies including comparison with both experimental data
      and other generators (KoralZ).
\item Processes of the type 
      $\mu^+\mu^- \to \gamma^*,Z^*,h^* \to f\bar f$ at higher energies
      including the effect of propagating Higgs bosons, both from the
      Standard Model and its minimal supersymmetric extension (MSSM). 
\end{itemize}

\subsection*{$e^+e^- \to f\bar f$}
The effect of ISR on processes of the type $e^+e^- \to f\bar f$
at LEP2 energies is tremendous. Due to the ``Radiative Return'' huge 
numbers of events actually reside on the $Z^0$ resonance instead of the 
beam energy.

The effect is twofold: First, the visible energy in the detector tends
to be reduced to energies of the order of the $Z$ mass accompanied by
an increase of the total cross section. Secondly, the 
forward--backward asymmetry due to the axial coupling of the $Z$ boson
tends to be shifted correspondingly.

We will highlight this effect in a series of figures. 
Fig.\,\ref{p_sigma189} exhibits the total cross section for
$e^+e^-\to \mu^+\mu^-(\gamma)$ in comparison with data from the DELPHI
collaboration \cite{delphi}.
Fig.\,\ref{p_asym189} depicts the combined forward--backward asymmetry
in $\mu$ and $\tau$ pair production at LEP2 compared to data taken 
by OPAL \cite{opal}.
Finally, Fig.\,\ref{p_costh189} shows the differential cross section with 
respect to the cosine of the angle between equally charged incoming
and outgoing leptons compared to data from L3 \cite{l3}.
To guide the eye, we have included the uncorrected Born level
prediction with a black line in the first two figures.

In all three cases different cuts on $s'$ have been applied, and in all 
cases the agreement of our generator with data is very good.
Note that we did not take into account any FSR effects.

\DOUBLEFIGURE{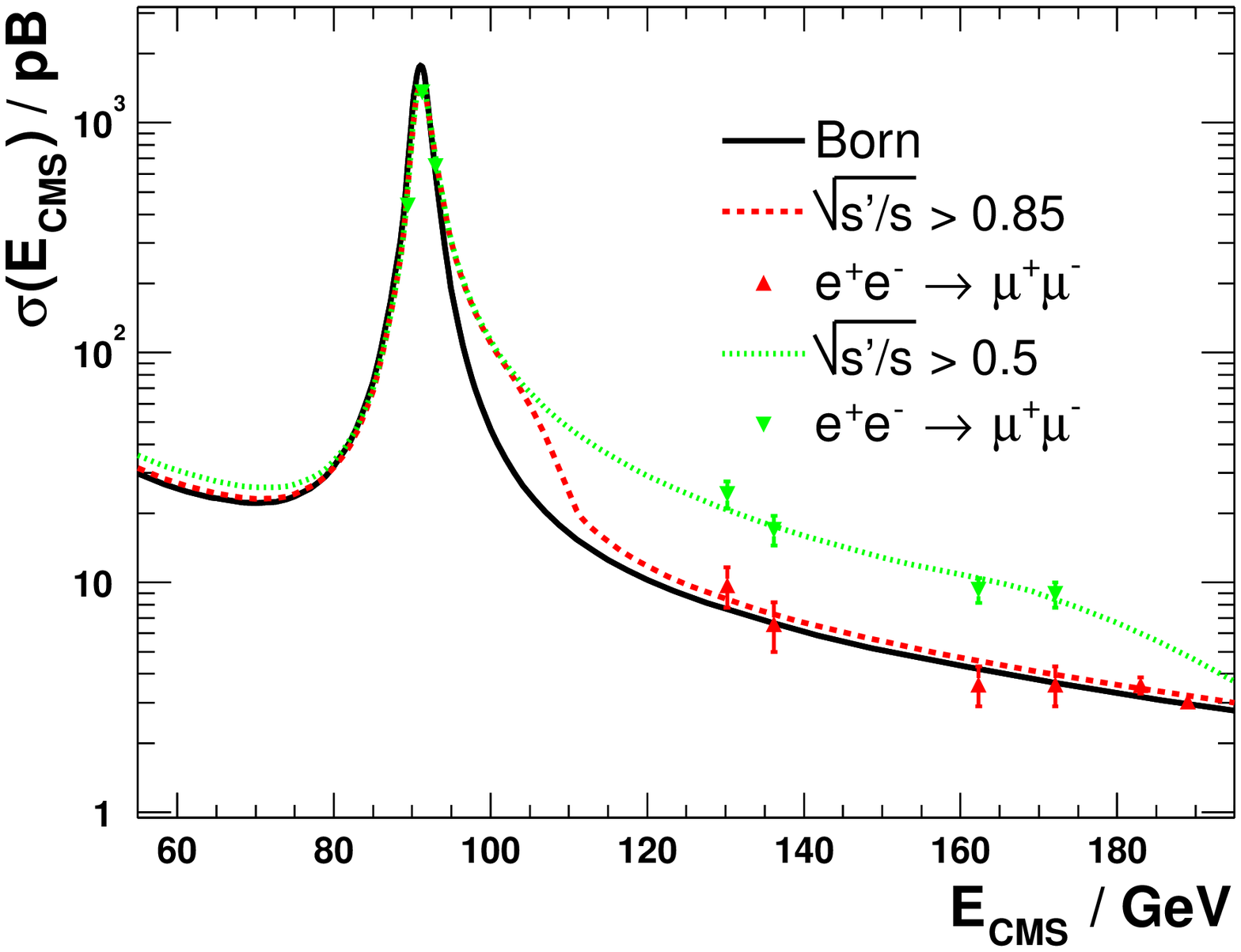,width=7.5cm}
{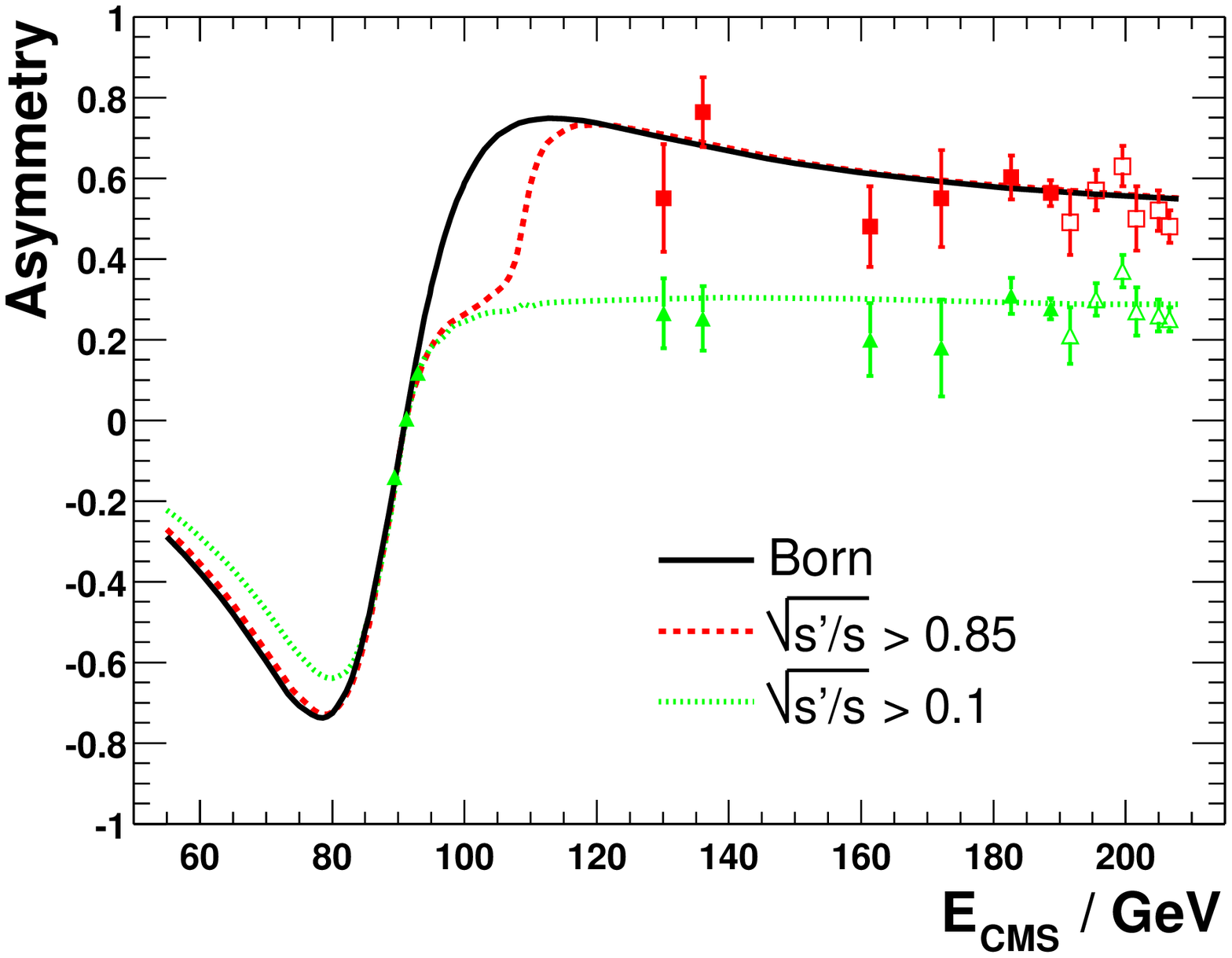,width=7.5cm}
{\label{p_sigma189}  Energy dependence of the cross section $\sigma$ at 
energies above the $Z_0$ boson mass in comparison with DELPHI data
\cite{delphi}. Two different cuts on $s'$ are applied, including and
excluding a radiative return to the $Z^0$ pole.}
{\label{p_asym189}  Forward--backward asymmetry above the resonance of the 
$Z_0$ boson in comparison with combined muon and tau pair data from
OPAL \cite{opal}. Again, different cuts on the effective c.m.\ energy
$\sqrt{s'}$ are plotted together with the Born expectation (without
ISR). The open symbols mark preliminary data \cite{opal_prel}.}

\EPSFIGURE{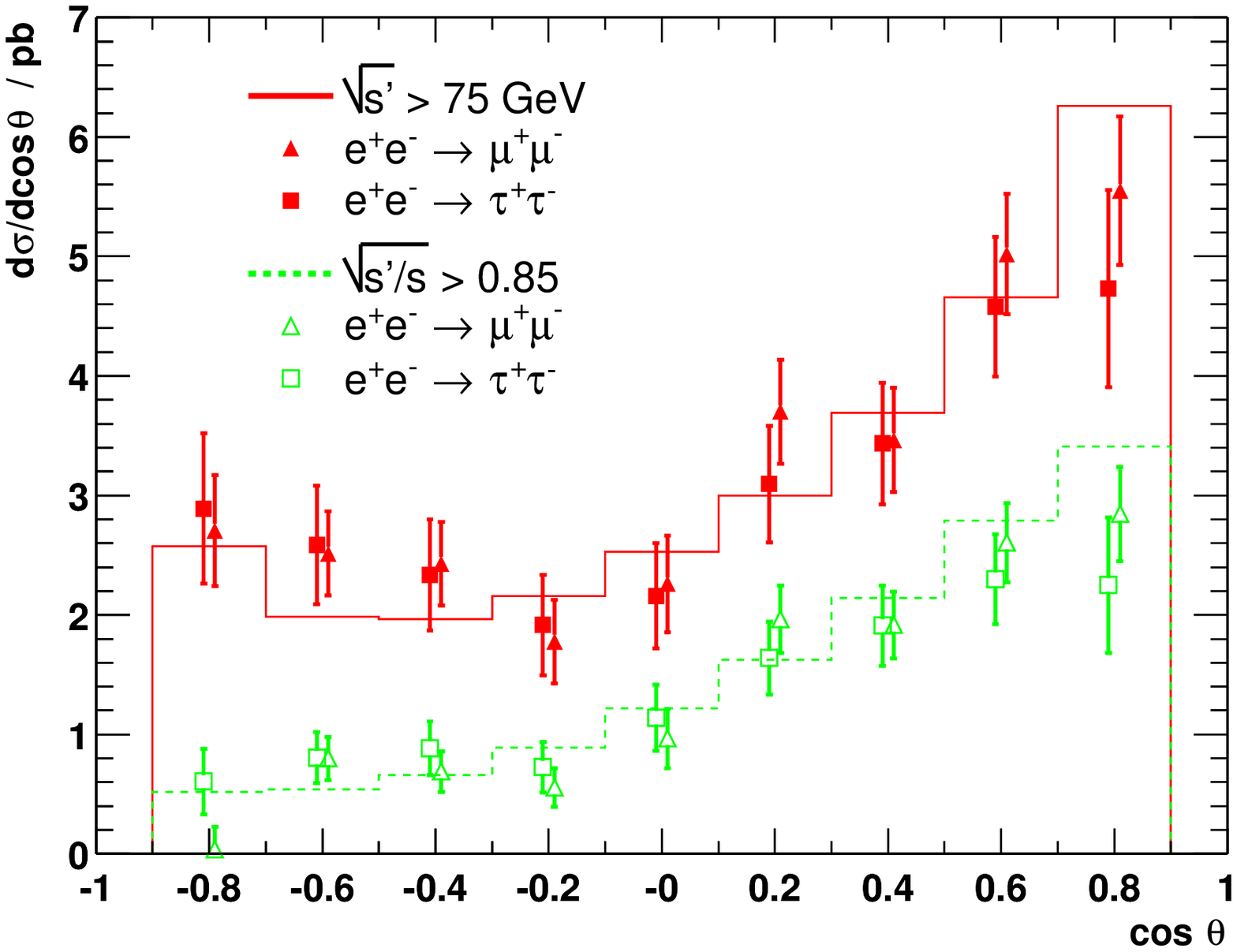,width=11cm}
{\label{p_costh189}  Differential cross section at $\sqrt{s}=189$ GeV in 
comparison with L3 data \cite{l3}. Different $s'$ cuts have been applied.}

In Figs.\,\ref{p_sprime189} and \ref{p_ephot189} the distribution of the 
effective c.m.\ energy $\sqrt{s'}$ of the $\mu^+ \mu^-$ pair and the 
distribution of photon energies are shown in comparison with results 
from KoralZ (second order). We find remarkably good agreement over 
almost the whole energy range in the $\sqrt{s'}$ distribution. The 
energy distribution is dominated by the contribution
of the hardest photon which is equally well described. The huge peak 
at 72.5 GeV corresponds to the radiation of one photon to reach the 
$Z$--resonance. Similarly, the edges at 48.9 GeV and 36.25 GeV 
correspond to a $Z$ return via emissions of two hard photons.
In this region our result differs slightly from KoralZ, since 
the interference effects become more pronounced. 
However, since the first order matrix element is used for 
the correction of every explicitly generated photon (not only the
hardest) some of the higher order matrix element effects are already
included. Note that we have concentrated on ISR effects only and
therefore switched off FSR and weak corrections in KoralZ. In case
only the first order correction of KoralZ is used, the results would
be nearly indistinguishable. 

\DOUBLEFIGURE{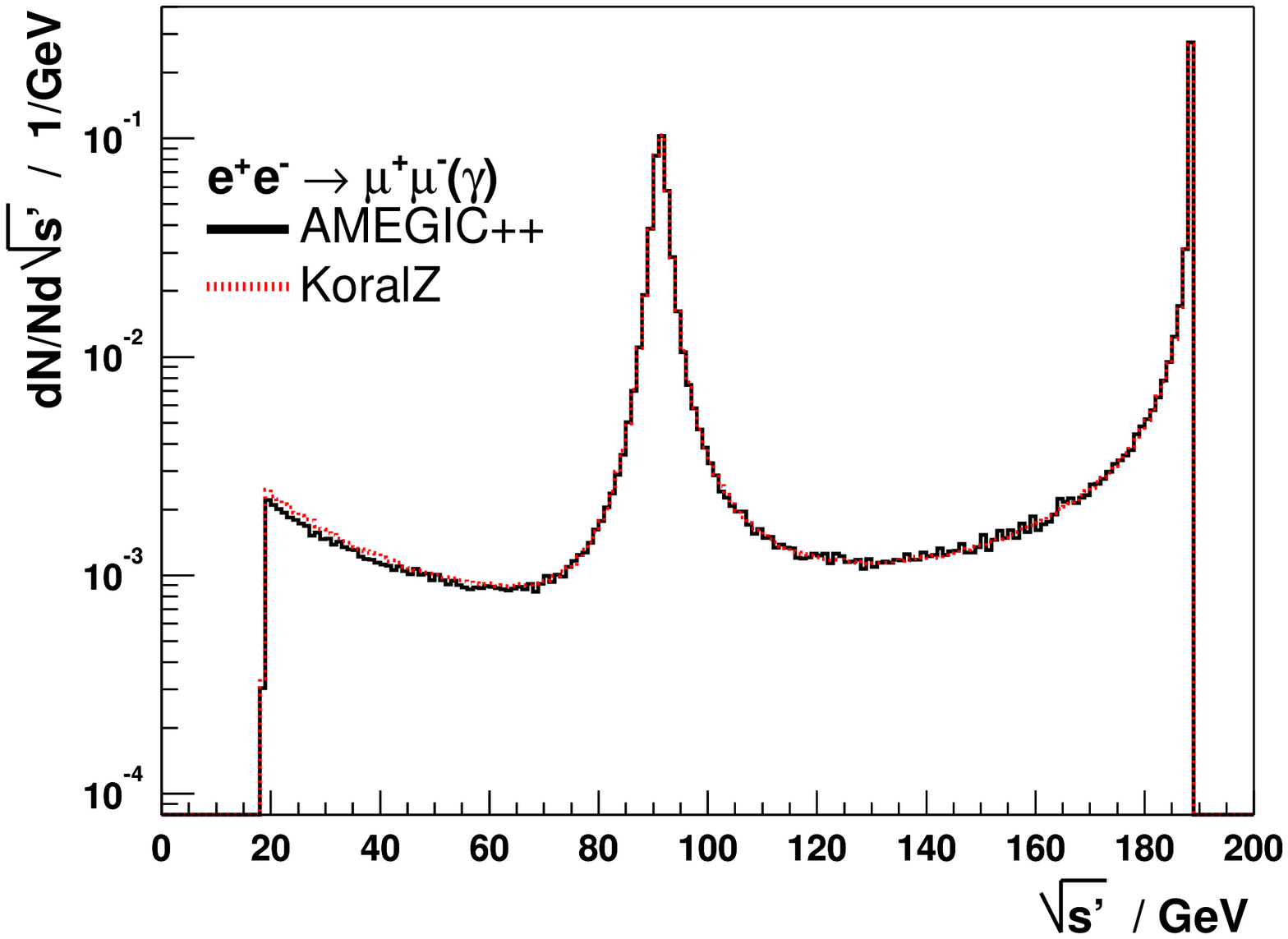,width=7.5cm}
{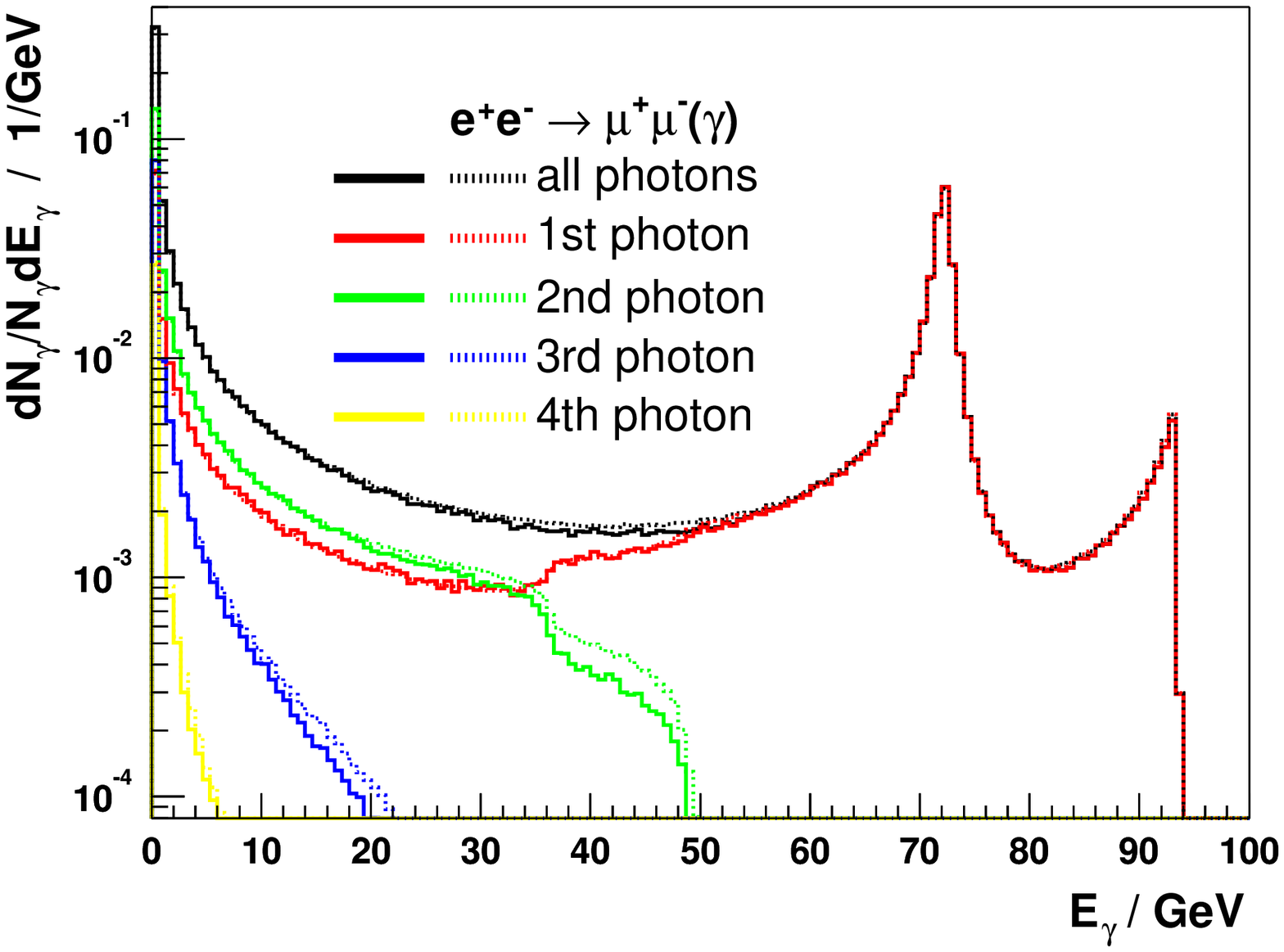,width=7.5cm}
{\label{p_sprime189} Distribution of the effective c.m.\ energy
$\sqrt{s'}$ of muon pairs at a beam energy of $\sqrt{s}=189\,{\rm
GeV}$. First order results by \AME\ are compared with the second order
result of KoralZ.} 
{\label{p_ephot189} Energy distribution of ISR photons in comparison
with KoralZ. The photons are ordered by their energies. Again, solid
lines are results produced by \AME, dotted lines are results from
KoralZ.}

\subsection*{Muon Collider}

As another application we consider a possible muon collider
\cite{muoncollider}. Its main purpose will be the precise measurement
of the properties of the Higgs boson once it is found. The muons are
used for two reasons: First, lepton colliders provide comparably pure
initial states allowing for spectroscopical measurements (like for
instance the exact determination of branching ratios of rare decays
like $h\to \gamma\gamma$). Second, muons have a non-negligible
coupling to the Higgs boson which in turn will be produced as an
$s$--channel resonance. 

In Fig.\,\ref{p_sigma115} and Fig.\,\ref{p_asym115} the total cross
section and the forward--backward asymmetry for 
$\mu^+ \mu^- \to b \bar b$ are plotted with and without ISR effects
assuming a Standard Model Higgs boson with a mass of 115 GeV. The
small width of roughly 4 MeV is an experimental challenge, e.g.\
effects of beam resolution have to be considered. The main effect of
ISR is to increase the background due to radiative return events to
the $Z^0$ peak. However, this background can be reduced via an
appropriate $s'$ cut. Of course, a radiative return to the Higgs peak
is also possible, but since the cross section is tiny the effect is
small and only visible close to the peak.

\DOUBLEFIGURE{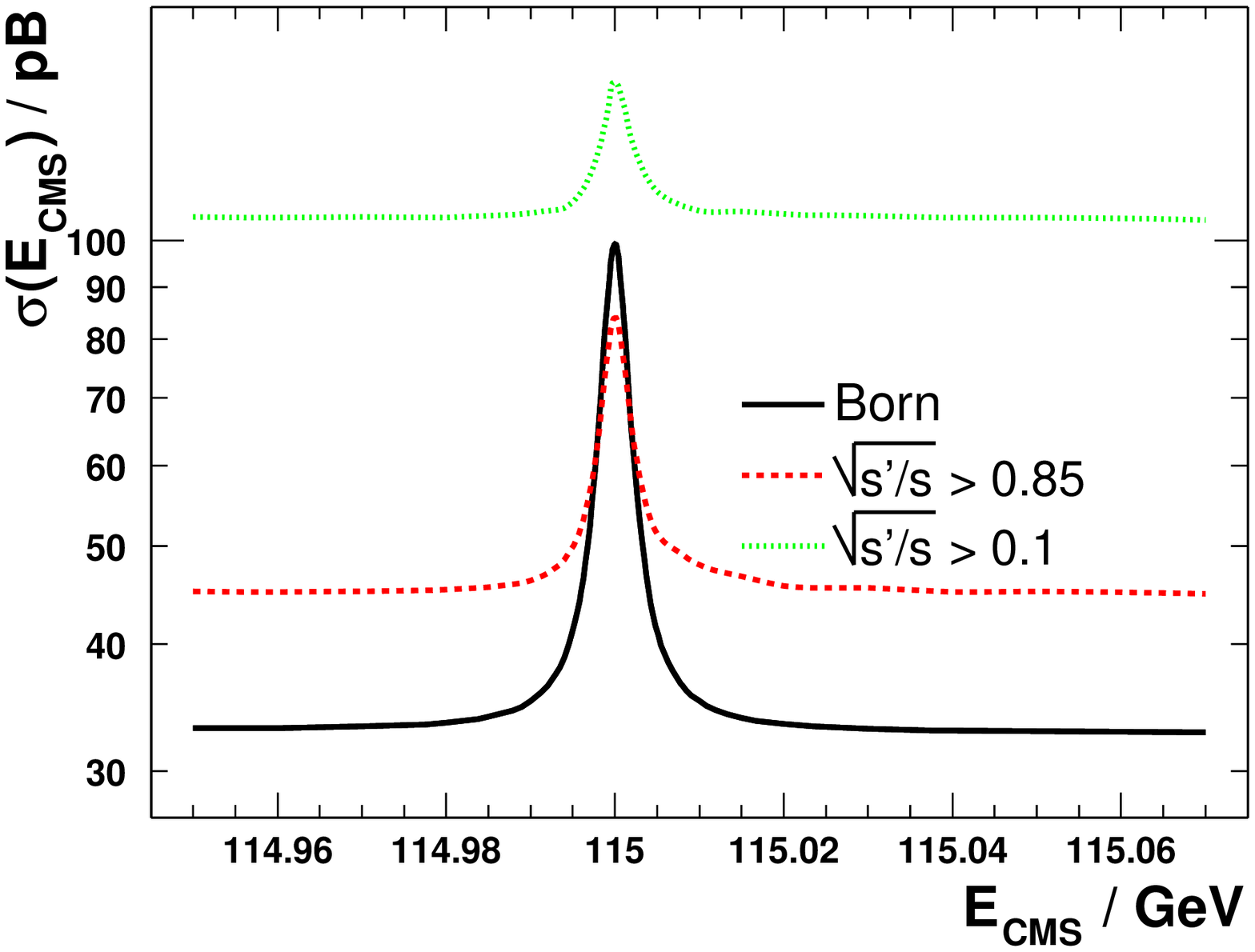,width=7.5cm}
{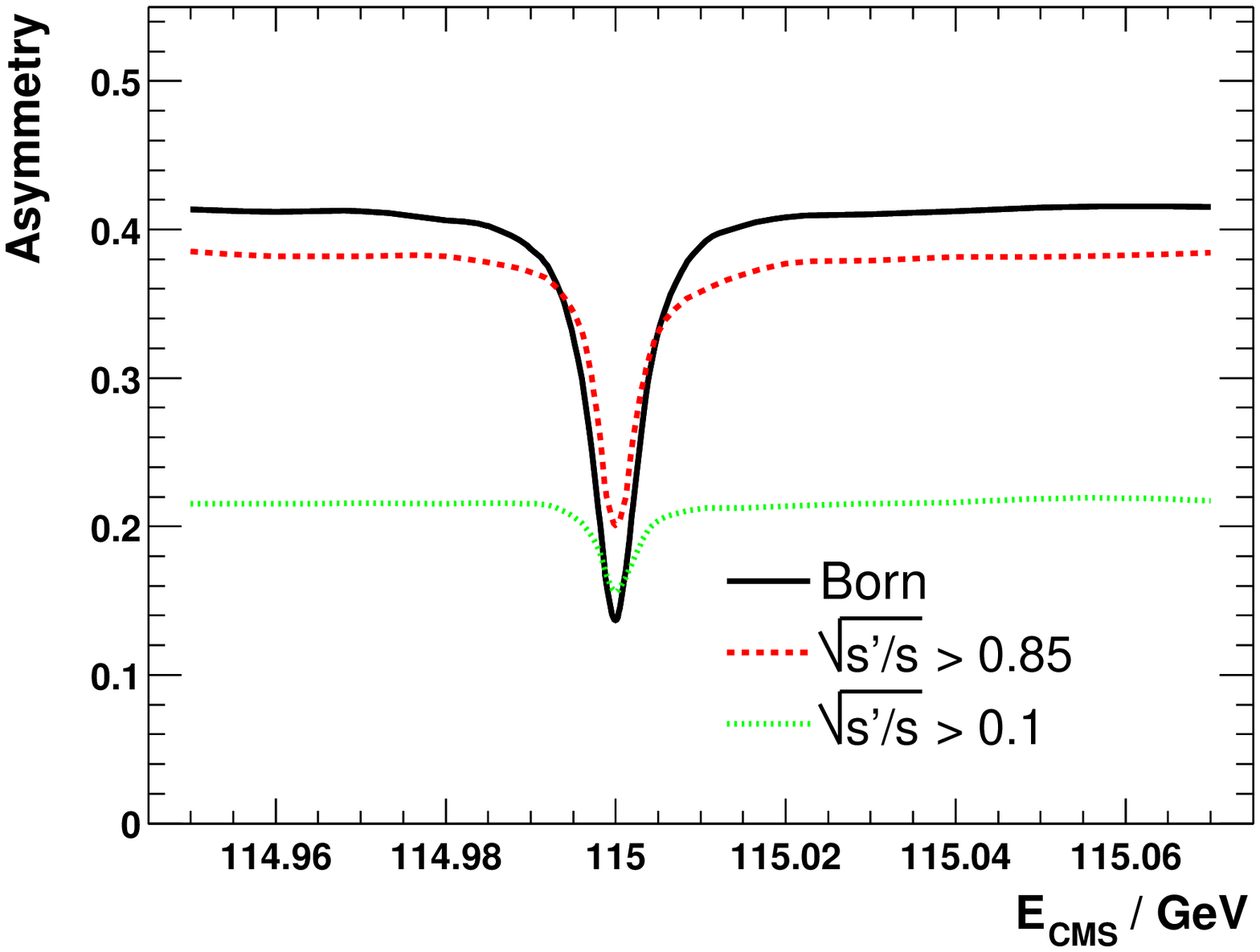,width=7.5cm}
{\label{p_sigma115}  Energy dependence of the cross section $\sigma$
around the resonance of the $h_{\rm SM}$ boson in the process 
$\mu^+\mu^- \to b \bar b$.} 
{\label{p_asym115}  Forward--backward asymmetry around the resonance
of the $h_{\rm SM}$ boson in the process $\mu^+ \mu^- \to b \bar b$.} 

Since our method is general, we can discuss further the extension to
any scalar boson propagating in the $s$--channel. As an example we
consider the case of Higgs bosons in the Minimal Supersymmetric
extension of the Standard Model (MSSM) \cite{MSSM,higgs_schannel}. In
this model there are two more neutral Higgs bosons which tend to be
considerably heavier than the light one. For the following discussion
let us constrain ourselves to a supergravity inspired supersymmetry
breaking scenario \cite{mSUGRA, point_sps1} with parameters to be
found in the Appendix \ref{a_parameters}. In this scenario we have a
mass of $m_{H_0} = 395.0$~GeV for the CP--even and and a mass of
$m_{A_0} = 394.5$~GeV for the CP--odd Higgs boson. The total cross
section can be found in Fig.\,\ref{p_sigma394} and the influence on
the forward--backward asymmetry is given in Fig.\,\ref{p_asym394}. At
this point in the mSUGRA parameter space the individual peaks can not
be resolved and the two Higgs bosons appear as one single, comparably
broad, resonance only. The effect of ISR is a slight shift of the peak
to larger c.m.\ energies and an increase of background due to
radiative return to the lower lying $Z$ boson resonance. However, the
latter effect can again be reduced by a suitable cut on the effective
c.m.\ energy $s'$. 

In striking contrast to the case of the lighter SM Higgs boson, 
in this case the radiative return to the peak structure from higher 
energies is important, see Figs.\,\ref{p_sprime394} and
\ref{p_sprime394d}. Note that the contribution of a photon, a $Z^0$
boson, and all three Higgs bosons propagating in the $s$--channel as
well as all interference terms have been taken into account. The
effect of the interplay of the different terms becomes apparent in
angular distributions, cf. Fig.\,\ref{p_costh394}, where different
cuts on the c.m.\ energy have been applied. The Born term is dominated
by the off--peak $Z$ boson contribution and exhibits therefore a
slight angular dependence. When taking the radiative return to the
heavy Higgs bosons peak into account, the distribution is only shifted
due to the angular blind scalar coupling structure. The predominating
contribution of the return to the $Z$ resonance restores the angular
distribution already known from Fig.\,\ref{p_costh189}.  

\DOUBLEFIGURE{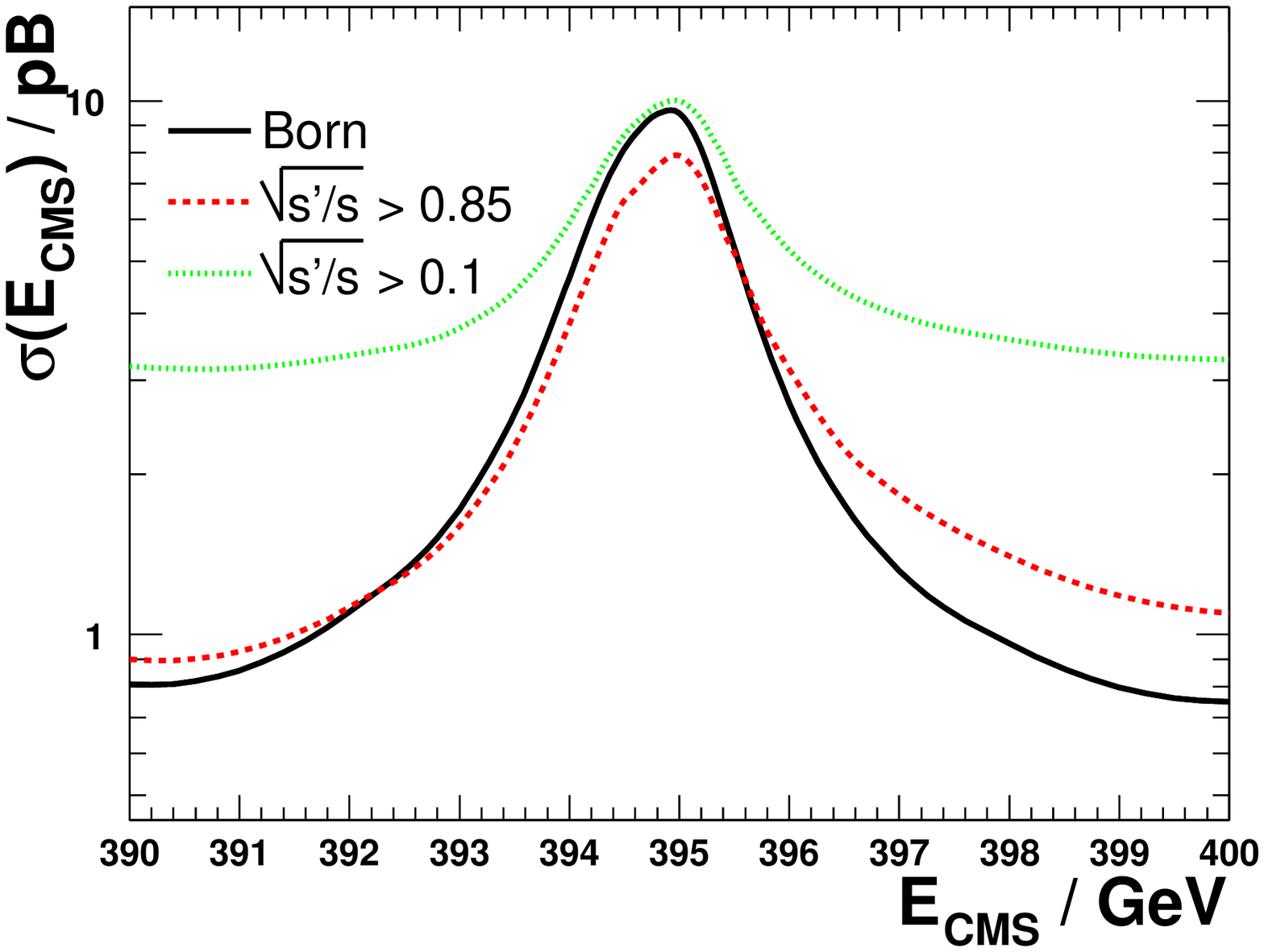,width=7.5cm}
{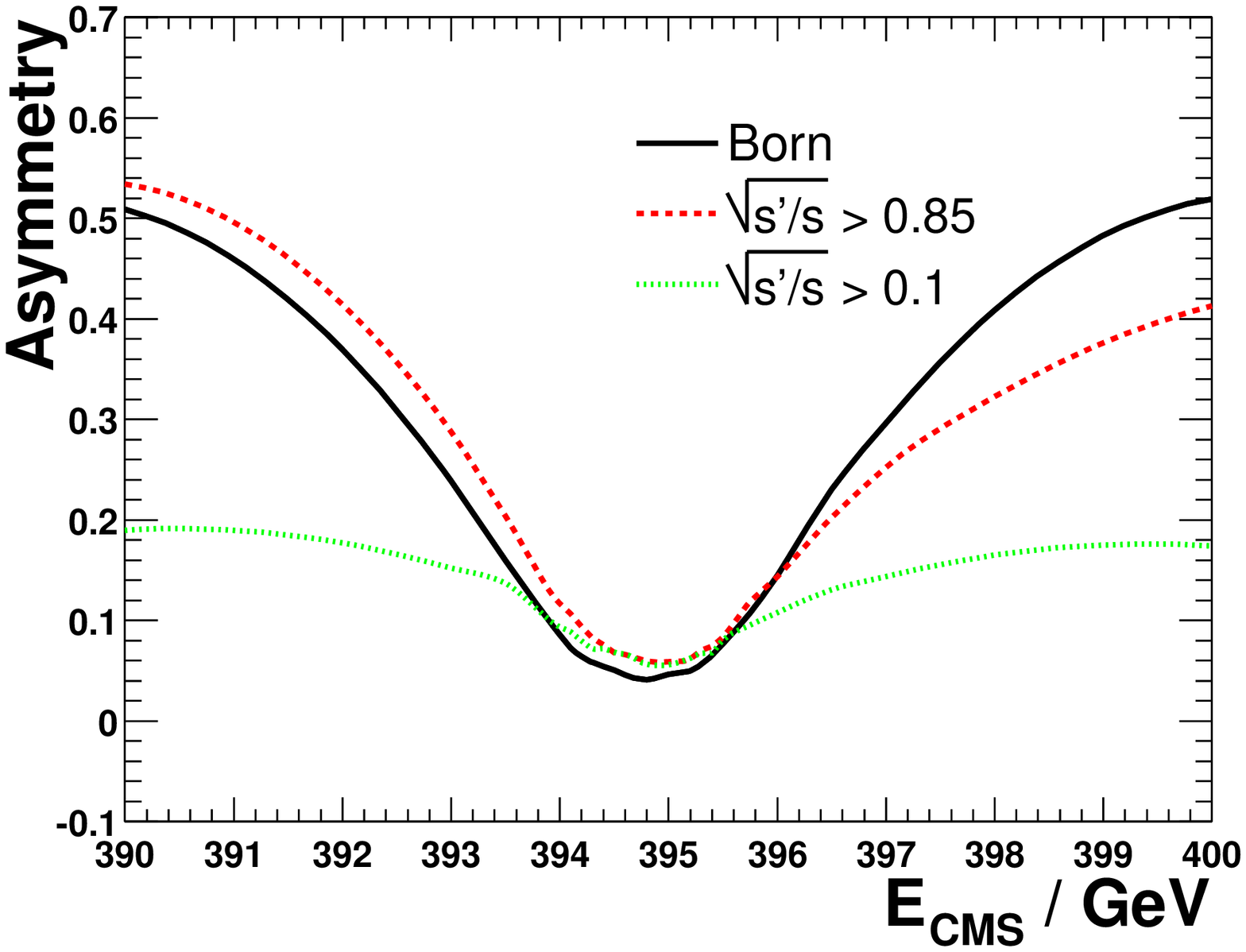,width=7.5cm}
{\label{p_sigma394} Energy dependence of the cross section $\sigma$
around the resonance of the heavy scalar Higgs boson $H_0$ and the
pseudo scalar Higgs boson $A_0$ in the process  
$\mu^+ \mu^- \to b \bar b$ for different $s'$ cuts, as well as without
ISR.} 
{\label{p_asym394} Asymmetry around the resonance of the heavy scalar
Higgs boson $H_0$ and the pseudo scalar Higgs boson $A_0$ with and
without ISR.} 

\DOUBLEFIGURE{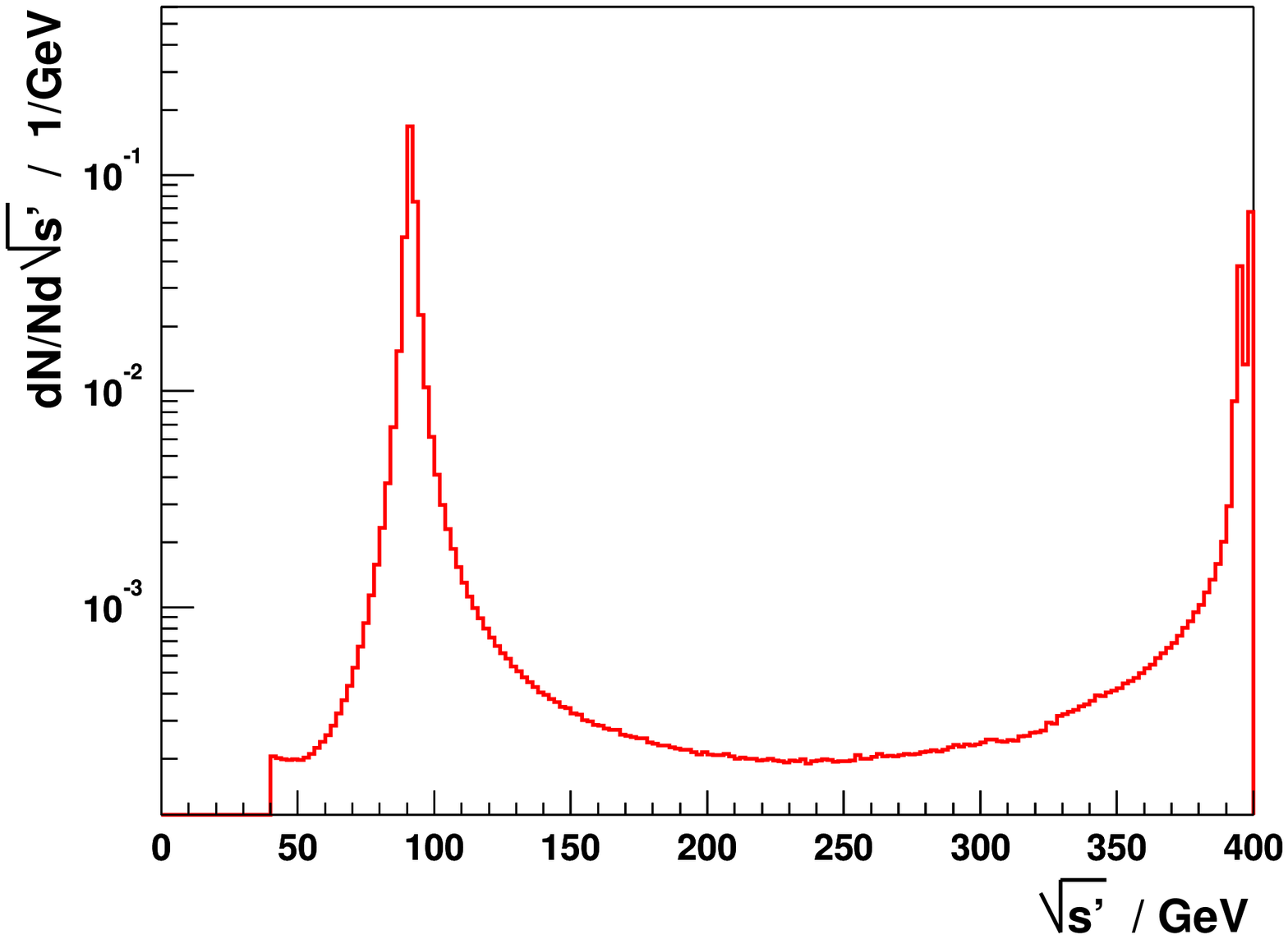,width=7.5cm}
{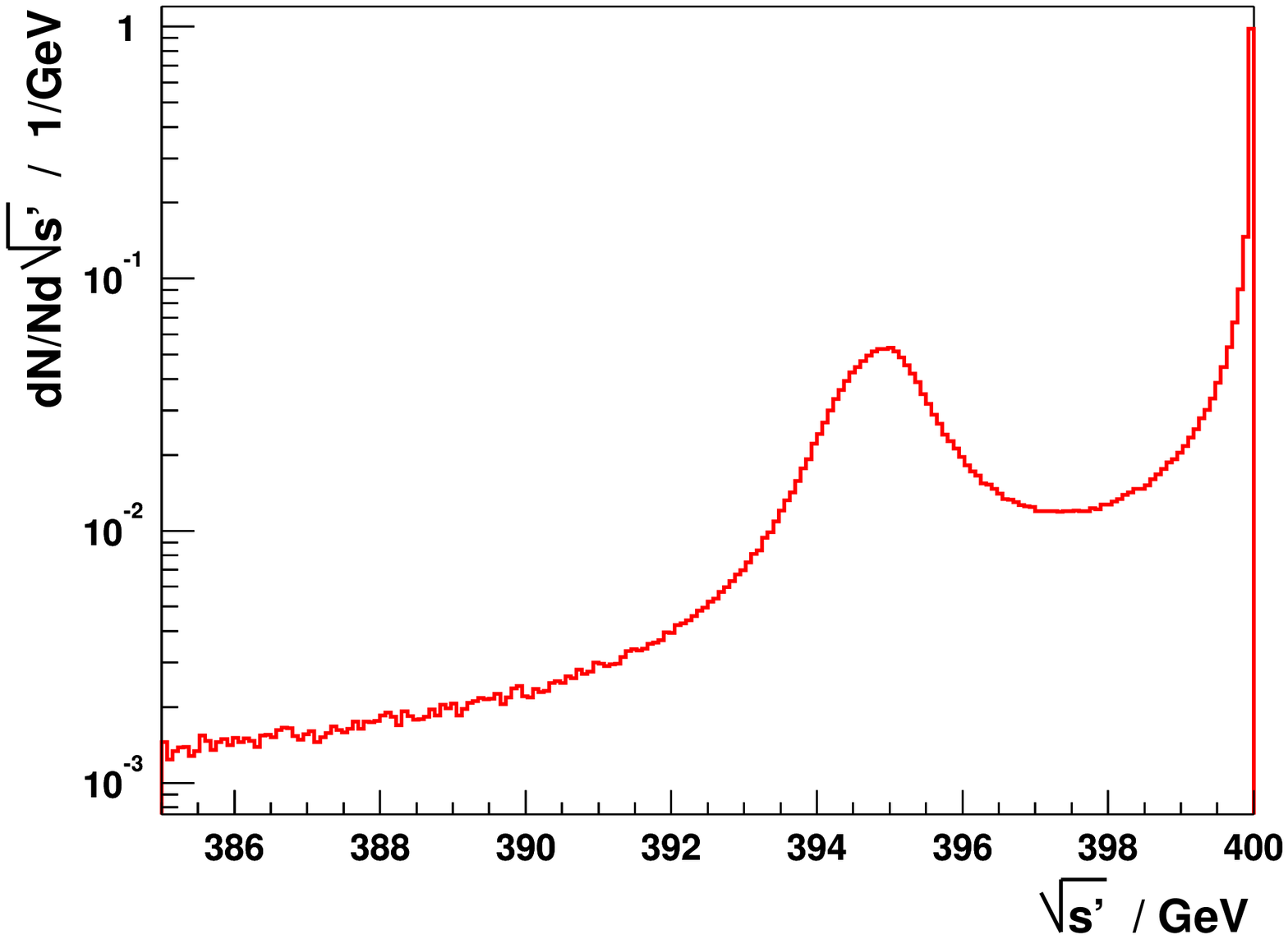,width=7.5cm}
{\label{p_sprime394}  Distribution of the effective c.m.\ energy
$\sqrt{s'}$ for the process  $\mu^+ \mu^- \to b \bar b$ at a beam
energy of $E_{\rm CMS}$=400 GeV.} 
{\label{p_sprime394d} Detail of Fig.\,\ref{p_sprime394}.}

\EPSFIGURE{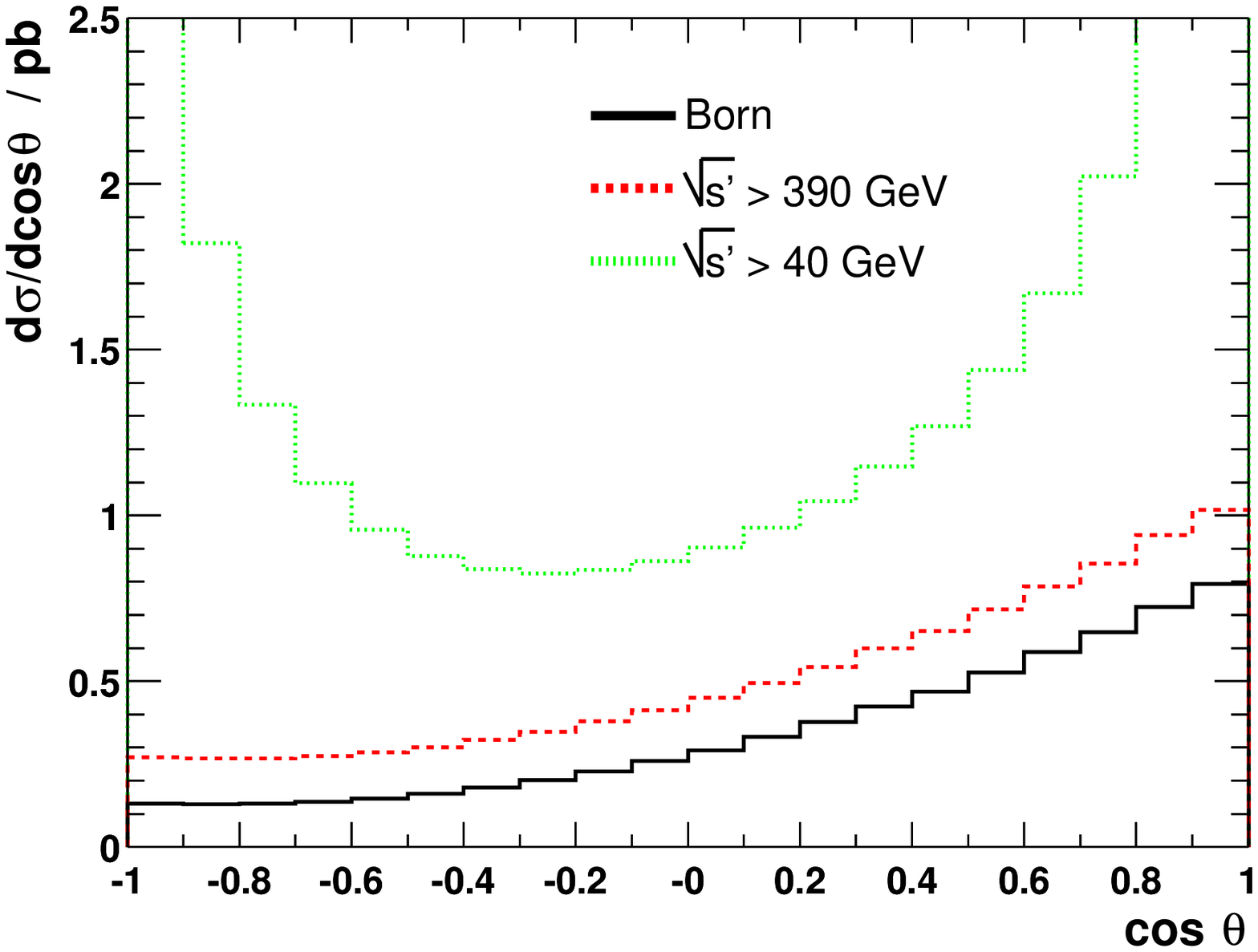,width=9cm} 
{ \label{p_costh394} Angular distribution of the b-quark with respect
to the incoming muon. Different cuts applied on $\sqrt{s'}$ in order
to include return events to the $Z^0$ or to the heavy Higgs bosons $H_0$
and $A_0$ only. For comparison the Born expectation is plotted too.}

\section{Conclusion and Outlook}
In this paper we presented an implementation of the successful YFS
algorithm into the matrix element generator \AME. To make full use of
the possibilities offered by automatic matrix element generation with
help of the helicity method and fully automatic integration of the
matrix element we developed a new method to deal with the first order
matrix element correction to multi photon emissions. We proved the
quality of our method by two examples: 
\begin{enumerate}
\item 
  For lepton pair production at energies at and above the $Z$ boson
  resonance, we have confronted our simulator with experimental data
  from the LEP experiment and another ISR generator, KoralZ. In both
  cases we found encouraging agreement. 
\item 
  The application of our approach to the $s$--channel production of
  Higgs bosons at a muon collider has illustrated the generality of
  our approach. Let us note that extensions to other bosons in the
  $s$--channel with other couplings is straightforward and will be
  addressed in the near future. 
\item 
  Furthermore, the interface of \AME\ with the parton shower
  implemented in \APA\ \cite{APA}provides a powerful tool for the
  simulation of hadronic final states including the effect of both 
  ISR and the final state parton shower.
\end{enumerate}     

We feel obliged to comment briefly on possible improvements:
\begin{itemize}
\item 
  Definitely, effects of Final State Radiation and its interference with ISR
  are missing and should be implemented. Additionally, some types of
  one--loop corrections like QCD corrections in the final state,
  electroweak corrections or box diagrams have not been implemented
  yet
  \footnote{During the time of writing we became aware of
  \cite{Dittmaier2002}, where an ISR calculation for a Higgs boson
  production at a muon collider was considered. This was based on the
  structure function approach and already includes FSR,
  IF--interference and QCD effects. Their findings are consistent with
  our Monte Carlo results.}.
  Also, an extension to second order corrections should be 
  done to achieve the precision needed for a proper interpretation of
  the anticipated precision data.
\item Apart from $s$--channel processes, at $e^+e^-$ colliders 
  processes like $ZZ$ production play a crucial role and deserve a
  similar treatment. In the case of for instance $WW$ production an
  extension of our method is not a trivial task. 
\end{itemize}

\section*{Acknowledgements}
We would like to thank K. Hamacher, U. Flagmeyer, U. M{\"u}ller, A. Behrmann 
from the Delphi collaboration and P. Ward and D. Ward from the Opal
collaboration for pleasant conversation and great help with the
experimental data. F.K. gratefully acknowledges financial help of the
DAAD. Furthermore, A.S. thanks BMBF for funding.  

This work was supported in part by EU Fourth Framework Programme
``Training and Mobility of Researchers'', Network ``Quantum
Chromodynamics and the Deep Structure of Elementary Particles'',
contract FMRX-CT98-0194 (DG 12 - MIHT).

\begin{appendix}
\section{The MC algorithm}\label{a_mc_alg}
In this appendix we would like to discuss the Monte Carlo procedure
for the generation of ISR photons, further details of the first three steps  
can be found in \cite{yfs2}.

\begin{enumerate}
\item \label{step_cms}Choice of the reduced c.m.\ energy squared

First, the reduced c.m.\ energy squared is determined. This is done 
according to a ``crude distribution'' that consists of a fast 
approximation for the energy dependence of the cross section. 
This approximation already exhibits the correct peak structure as 
given by the $s$--channel propagators. Additionally, it includes the 
leading logarithmic photon factor. The corresponding probability 
density in terms of the energy fraction $\nu = 1 - s'/s$ reads
\bea\label{mc_density}
f(\nu) = \frac{\gamma }{(1-\nu)\nu^{1-\gamma}}\,\Psi(\nu)
\eea
with    
\bea
\Psi(\nu) = (1-\nu)\,\sigma_{\rm Born}(s') \;
     \hat{\cal J}(\nu) \quad \mbox{and} \quad 
\gamma =  \frac{2\alpha}{\pi}\left(\ln\frac{s}{{m}^2}-1\right)\ln\eps\;.
\eea
The energy fraction $\nu$ can be determined according to $f(\nu)$  
by standard Monte Carlo sampling techniques. In the equation above 
$\hat{\cal J}(\nu)$ is an approximate form of the Jacobian that will
be discussed in step 3(b).
\item Choice of number of resolvable photons

Having determined the effective center of mass energy squared 
$s'=s(1-\nu)$, we proceed by choosing the number of resolvable
photons which in turn have to be generated explicitly. In the logarithmic 
approximation the photon number $N_\gamma$ follows a shifted Poisson
distribution: 
\be
f(N_\gamma-1) = e^{-\mu}\frac{\mu^{N_\gamma-1}}{(N_\gamma-1)!}\;.
\ee
Obviously, its mean value depends on both $s'$ and the photon resolution 
parameter $\eps$. It is given by
\bea
\mu & =& \int\frac{d^3 \bvec{k}}{k^0}
     \St(p_1,p_2,k)
     \Theta\left(\frac{2 k^0}{\sqrt{s}}-\eps\right)
     \Theta\left(\nu-\frac{2 k^0}{\sqrt{s}}\right)  \\
  & = &
 \frac{2\alpha}{\pi}
 \ln\left(\frac{s}{m_e}\right) \ln\left(\frac{\nu}{\eps}\right)
 \Theta\left(\nu-\eps\right)
 \;.
\eea
This can be identified in the master formula, Eq.(\ref{master_f}), 
considering the additional constraint on the photon energies 
(the second $\Theta$--function) not to exceed $s'$.
\item Construction of four--momenta

The construction of four--momenta for the photons is achieved
in two steps:
\begin{enumerate}
\item Angles: 
The photon angles are now individually appointed to each photon with 
the help of the eikonal $\St$. Clearly, $\St$ exhibits no dependence on
the azimuthal angle $\varphi$, hence this angle is distributed
uniformly. In contrast, $\St$ depends strongly on the polar angle 
$\vartheta$, which in turn is chosen according to
\be\label{cos_dist}
\bar{f}(\cos\vartheta) =
     \frac{1}{(1-\beta\cos\vartheta)(1+\beta\cos\vartheta)}
     \quad \mbox{with} \quad
     \beta = \sqrt{1- \frac{4 m^2}{s}}\;.
\ee
Note that terms proportional to the lepton mass have been discarded at this 
stage. Therefore, a correction weight is introduced for each photon
\be
  w_{\rm angle}(\vartheta) = \frac{f(\vartheta)}{\bar{f}(\vartheta)}\;,
\ee
where the exact distribution $f(\vartheta)$ is given by
\be
 f(\vartheta) =  
   \bar{f}(\cos\vartheta) - \frac{m^2}{s-2 m^2} 
   \left( \frac{1}{(1-\beta\cos\vartheta)^2} +
          \frac{1}{(1+\beta\cos\vartheta)^2} \right)\;. 
\ee
Thus, the weight to reproduce the correct mass--dependence reads
\be
  w_{\rm mass} = \prod_{i=1}^{N_{\gamma}} {w_{\rm angle}(\vartheta_i)}
 \;.
\ee
\item Energies: 
The next step is the assignment of energies to each of the photons. In 
case that exactly one photon above the threshold has been generated, the 
energy of this photon is already completely constrained by $\nu$ and the 
chosen photon angles due to overall four--momentum conservation. 
In the case that more than one photon have been generated, the
additional photon energies are chosen according to $dk^0/k^0$, i.e.\
the eikonal $\St$. Then a scaling of all photons is performed in order
to achieve four--momentum conservation. Thereby an Jacobian 
\be
{\cal J}({K},\nu)  = 
\half \left( 1 + \frac{1}{\sqrt{1-A\nu}} \right)
\label{f_jac}
\ee
with the abbreviations
\be
A  = \frac{{K}^2 P^2}{({K}P)^2}\;,\quad K = \sum_{n=1}^{N_\gamma} k_n
\ee
is introduced. It had been taken into account already in step
\ref{step_cms} as $\hat{\cal J}$ assuming $A=1$, since the photon
momenta have not been know at this time. A correction weight 
\be
  w_{\rm jac}=\frac{{\cal J}({K},\nu)}{\hat{\cal J}(\nu)}
\ee
is introduced to cure this simplification. Furthermore, all photon
energies have to be above the limit $\eps \sqrt s / 2$ after
rescaling, in order to avoid double counting. This leads to an
additional weight 
\be
  w_\eps=\Theta\left(k_n^0 -  \frac{\sqrt{s}}{2} \eps \right)\,.
\ee
\end{enumerate}
\item Matrix element correction

The next step is the calculation of the matrix element correction. This is 
done by introducing an additional weight according to
\begin{equation}
  w_{\rm ME}  =     \frac{1}{\sigma_0(s')}
  \biggl\{  \beta^{(1)}_0(p_i,q_j) + \sum_{n=1}^{N_\gamma}
  \frac{\beta_1^{(1)}(p_i,q_j,k_n)}
                         {\St(p_i,k_n)} 
  \biggr\}
  \;.
\end{equation}
\item Re-weighting

The last step is the calculation of the final weight in order to cure
the approximations made during the choice of the reduced c.m.\ energy
(step 1) and the determination of the photon momenta (step 3) as well
as to incorporate the matrix element corrections 
\begin{equation}
  w_{\rm MC} = w_{\rm mass} w_{\rm jac} w_{\eps} w_{\rm ME}
  \;.
\end{equation}
This weight can now be used in a rejection method to produce
unweighted events.  
\end{enumerate}

\noindent
The ISR simulation in \AME\ differs from this general description 
mainly in steps 1 and  4.

In step 1, a multichannel method is used to determine $s'$ to incorporate 
complicated peak structures. It can therefore be understood as the first part 
of the multichannel determination of the final state momenta. A library of
fast $f\bar f \to f' \bar f'$ cross sections for arbitrary scalar and
vector $s$--channel resonances enables the simulation
of many processes in the SM and beyond. Due to the rejection mechanism
these cross sections can be also used for multi-jet productions.

In step 4 the evaluation of the matrix element correction is
modified as described in section \ref{c_method}, thereby taking
full advantage of the features of \AME.

\section{Program Parameters}\label{a_parameters}

\DOUBLETABLE{
\begin{tabular}{|rl|}
\hline
$\alpha_{\rm QED}$    &  1./137.035995 \\
$\sin^2 \vartheta_w$  &  0.23117 \\
$M_Z$                 &  91.1882  GeV \\
$\Gamma_Z$            &  2.4952 GeV \\
$m_b$                 &  5.0 GeV \\
$m_c$                 &  1.3 GeV \\
$m_\mu$               &  105.658 MeV \\
$m_\tau$              &  1.777 GeV \\
\hline
$v$                   &  246.2 GeV \\
$m_{h_{sm}}$          &  115.0 GeV \\
$\Gamma_{h_{sm}}$     &    3.8 MeV \\
\hline
\end{tabular}
}
{
\begin{tabular}{|rl||rl|}
\hline
$M_0$          & 100. GeV     & $m_{H_0}$      & 395.0 GeV\\
$M_{1/2}$      & 250. GeV     & $\Gamma_{H_0}$ & 1.07 GeV\\
$A_0$          & -100. GeV    & $m_{A_0}$      & 394.5 GeV\\
$\tan\beta$    & 10.          & $\Gamma_{A_0}$ & 1.29 GeV\\
$\mu$          & +1           &  & \\
$m_t$          & 175. GeV     &  &\\
\hline
\end{tabular}
}
{\label{t_sm_para}Standard Model parameter used for the simulation.}
{\label{t_mssm_para}The mSugra point SPS1 according to \cite{point_sps1} 
 and the corresponding properties of the heavy Higgs bosons in the MSSM. The 
 widths of the Higgs bosons have been calculated  with \AME. }

The parameters used for the calculation of results presented in
section \ref{results} are given in Tab.\,\ref{t_sm_para} and
Tab.\,\ref{t_mssm_para}. 

To include higher order effects in the Yukawa coupling we have used a
running b--quark mass according to the well known relation
\be
  m_{\mu} = m_{\mu_0} 
            \left( 
            \frac{ \alpha_s(\mu)}{\alpha_s{\mu_0}}
            \right)^{\frac{\gamma_{m,0}}{\beta_0}}  
\ee
with 
\be
  \gamma_{m,0}=1\;,\quad \beta_0=\frac{1}{12} ( 33 - 2  n_f) \;.
\ee

\end{appendix}

\end{document}